\documentclass[12pt,thmsa]{article}%
\usepackage{amssymb}
\usepackage{amsmath}
\usepackage{sw20aip}
\usepackage{amsfonts}
\usepackage{graphicx}%
\begin{document}

\title{ADS/CFT Applied\ To Vector Meson Emission From A Heavy Accelerated Nucleus}
\author{Pervez Hoodbhoy\\Department of Physics\\Quaid-e-Azam University\\Islamabad 45320, Pakistan. \\{~}}
\date{11 September 2008}
\maketitle

\begin{abstract}
We consider a classical source, moving on the 4-D boundary of a 5-D ADS space,
that is coupled to quantum fields residing in the bulk. Bremsstrahlung-like
radiation of the corresponding quanta is shown to occur and the S-matrix is
derived assuming that the source is sufficiently massive so that recoil
effects are negligible. As an illustrative example, using the ADS hard-wall
model, we consider vector mesons coupled to a heavy nucleus that is moved
around at high speed in an accelerator ring. The meson radiation rate is found
to be finite but small. Much higher accelerations, such as when a pair of
heavy ions suffer an ultra peripheral collision, cause substantial emission of
various excited vector mesons. Predictions are made for the spectrum of this
radiation. A comparison is made against existing photon-pomeron fusion
calculations for the transverse momentum spectra of rho mesons. These have the
same overall shape as the recently measured transverse momentum distributions
at RHIC.

\end{abstract}

The gauge/gravity correspondence\cite{Theory} is a powerful means for
extracting information about four-dimensional strongly coupled gauge theories
by mapping them onto gravitational theories in five dimensions where, because
of the weak coupling, they may be solved much more easily. A highly prized
goal is to learn about non-perturbative QCD from some 5-D theory. This goal is
still some distance away because the gravity theory actually dual to QCD is
not yet known. Nevertheless, using variants of the $N=4$ supersymmetric
models, there have been a large number of interesting applications . These
include hard QCD scattering and deep inelastic structure functions\cite{DIS},
low lying hadron spectra\cite{Spectrum}, chiral symmetry breaking\cite{Son},
vector-meson couplings\cite{Rho}; meson form factors\cite{FF} moments of
generalized parton distribution \cite{GPD}, kaon decays\cite{Kaon}, etc. There
are many valid criticisms of the holographic approach\cite{Cohen} but, on the
whole, the reasonable agreement with experiment suggests that ADS ideas
deserve further exploration.

This work aims at extending the range of problems to which ADS ideas have been
applied. Since this is an illustrative calculation, for simplicity we shall
use the well-known hard-wall model. This uses an abrupt cutoff in ADS space.
While unsatisfactory in describing meson Regge trajectories, it is the
simplest way of enforcing confinement in this "bottom-up" approach. However,
it should be possible to generalize the contents of this paper to the
"soft-wall", designed to give the correct Regge behaviour \cite{Son}. We
choose vector fields since they have the simplest ADS description.

Let us quickly review the standard ADS approach to QCD: the gravity theory is
defined on a (d+1)-dimensional Anti-de Sitter $AdS_{d+1}$ space with a
$d$-dimensional asymptotic boundary at $z=0.$ The fields $\Phi(z,x)$%
\ propagate in $AdS_{d+1}$ and approach\ \ the conformal field theory (CFT)
fields $\Phi_{0}(x)$ on the boundary$.$ Various QCD composite operators
$J(x)$, which are built from quark and gluon operators and exist only on
$z=0$, act as sources for $\Phi(z,x).$\ They essentially serve as mathematical
devices by which to probe the bulk. On the gravity side the generating
functional is,%
\[
Z_{grav}=e^{iS_{eff}[\Phi]}=\int D\Phi e^{i\Phi},
\]
while on the CFT side, in the presence of the operator probes $J(x)$,%

\[
Z_{CFT}=e^{iS_{CFT}[\Phi_{0}]}=\int D\Phi e^{iS_{QCD}+i\int d^{d}xJ(x)\Phi
_{0}(x)}.
\]

The duality between the physics on the boundary and in the bulk is then
succinctly expressed by the equality,
\[
Z_{grav}[\Phi\rightarrow\Phi_{0}]=Z_{CFT}[\Phi_{0}]
\]

In the supergravity approximation, $Z_{grav}$ is easily calculated. Functional
differentiation with respect to $J(x)$ yields the desired correlation
functions of fields such as $\left\langle \Phi_{0}(x)\Phi_{0}(x^{\prime
})\right\rangle .$ With $J(x)$ having served its purpose, it can be set equal
to zero.

The approach taken here will be slightly different. We shall take $J(x)$ to be
an isovector source that excites fields in the bulk with the right quantum
numbers. However it will be a "real" source, not a fictitious one. This is
analogous to a time varying electrical current that couples to the
electromagnetic field and radiates photons. Provided that the energy radiated
is small, the recoil is negligible. Similarly, we shall assume that the
back-reaction on the iso-vector source radiating vector mesons can also be
ignored. The limitaions of this approach will be discussed.

\section{S-Matrix}

$\smallskip$With $R$ as the curvature of the AdS$_{5}$ space, the metric has
the conventional form,%
\begin{equation}
ds^{2}=\frac{R^{2}}{z^{2}}(\eta_{\mu\nu}dx^{\mu}dx^{\nu}-dz^{2}),\text{ for
}z_{0}>z>0.
\end{equation}
where $\eta_{\mu\nu}=(1,-1,-1,-1),$ $z=0$ is the 4-dimensional world sheet,
and $z_{0}$ is the distance at which the AdS$_{5}$ ends. The indices $\mu,\nu$
run from 0 to 3 while AdS$_{5}$ indices, denoted by $m,n$ run over 0, 1, 2, 3,
$z$. Limiting our attention to vector mesons, the bulk action is,%

\begin{align}
S  &  =-\frac{1}{4g_{5}^{2}}\int d^{5}xTr[F^{mn}F_{mn}]\\
F_{mn}  &  =\partial_{m}A_{n}-\partial_{n}A_{m}-i[A_{m},A_{n}].
\end{align}
\qquad

The field $A_{m}$ transforms under flavour $SU(N)$, $A_{m}=A_{m}^{a}t^{a}.$
Suppressing the flavour index, and with the gauge choice $A_{z}(z,x)=0,$ the
linearized equation of motion reads,
\begin{equation}
\lbrack z\partial_{z}(z^{-1}\partial_{z})-\square]A^{\mu}=0 \label{A1}%
\end{equation}
with $\square=\partial^{\mu}\partial_{\mu}$ being the usual D'Alembertian
operator. As the boundary condition, we require that $A^{\mu}(0,x)\ =0$ and
that $F_{mn}$ vanish at $z=z_{0}.$ The latter implies the Neuman boundary
condition, $\frac{\partial A^{\mu}}{\partial z}=0.$ Both conditions are
satisfied by $A^{\mu}=\varepsilon^{\mu}e^{-ik\cdot x}zJ_{1}(kz)$ for any $k$
that satisfies $J_{0}(kz_{0})=0.$ Since $k^{2}=k^{\mu}k_{\mu}=m^{2}$, this
implies a tower of vector mesons with masses given by $m_{p}=\frac{\chi_{p}%
}{z_{0}}$, where $J_{0}(\chi_{p})=0,\;p=1,2,\cdot\cdot\cdot$

Thus, the most general solution of Eq. \ref{A1} is,%
\begin{equation}
A^{\mu}(z,x)=\sum_{p=1,\lambda}^{\infty}\int\frac{d^{4}k}{(2\pi)^{4}}%
a_{p}(k,\lambda)\varepsilon^{\mu}(k,\lambda)e^{-ik\cdot x}zJ_{1}(kz)2\pi
\delta(k^{2}-m^{2})+cc. \label{field}%
\end{equation}
Canonical quantization now follows in a rather obvious way\cite{Braga} by
imposing the commutation relation,
\begin{equation}
\left[  a_{p}(k,\lambda),a_{p^{\prime}}^{\dagger}(k^{\prime},\lambda^{\prime
})\right]  =\delta_{pp^{\prime}}\delta_{\lambda\lambda^{\prime}}\delta
^{3}(k-k^{\prime}).
\end{equation}
This leads to,
\begin{align}
\left[  A^{\mu}(z,x),A^{\mu^{\prime}}(z^{\prime},x^{\prime})\right]   &
=zz^{\prime}\sum_{p=1}^{\infty}\frac{1}{z_{0}^{2}c_{p}^{2}}\triangle_{p}%
^{\mu\mu^{\prime}}(x-x^{\prime})J_{1}(k_{p}z)J_{1}(k_{p}z^{\prime
}),\label{comm}\\
\triangle_{p}^{\mu\mu^{\prime}}(x)  &  =\int d\widetilde{k_{p}}\left(
e^{-ik\cdot x}-e^{ik\cdot x}\right)  (-g^{\mu\mu^{\prime}}+\frac{k^{\mu}%
k^{\mu^{\prime}}}{k^{2}}),\\
c_{p}^{2}  &  =\int_{0}^{1}dyyJ_{1}^{2}(\chi_{p}y)\\
d\widetilde{k_{p}}  &  =\frac{d^{3}k}{(2\pi)^{3}2\omega_{p}}\text{ with
}\omega_{p}^{2}=\left\vert \overrightarrow{k}\right\vert ^{2}+m_{p}^{2}.
\end{align}
The above sum over momenta is restricted to discrete values, $k_{p}=\frac
{\chi_{p}}{z_{0}}$.

\ The field in the bulk arising from a source $J^{\mu}(x)$ placed on the $z=0$
boundary is,%
\begin{equation}
A^{\mu}(z,x)=g\int d^{4}xG(z,x-x^{\prime})J^{\mu}(x^{\prime}), \label{A2}%
\end{equation}
where the Green's function $G(z,x-x^{\prime})$\ is a sum of retarded and
advanced parts, $G=G_{R}+G_{A}$. It will be computed using the basis provided
by the solutions of Eq. \ref{A1}.

To this end, let us find solutions to
\begin{equation}
\lbrack z\partial_{z}(z^{-1}\partial_{z})-\square]G(z,x)=\delta^{4}%
(x-x^{\prime}).
\end{equation}
After Fourier transformation, the solution can be written as,%
\begin{equation}
G(z,x)=\int\frac{d^{4}k}{(2\pi)^{4}}e^{-ik\cdot x}zf(k,z),\text{ \ }
\label{A3}%
\end{equation}
where $f(k,z)$ obeys,\
\begin{equation}
\left[  z^{2}\frac{d^{2}}{dz^{2}}+z\frac{d}{dz}+(k^{2}z^{2}-1)\right]
f(k,z)=z. \label{A4}%
\end{equation}
With the boundary conditions $f(k,0)=0$ and $(zf)^{\prime}\!(z=z_{0})=0$ , Eq.
\ref{A4} yields a complete, orthogonal set of solutions $\left\{  J_{1}%
(k_{p}z)\right\}  $ which allow for the delta function expansion,
\begin{equation}
\frac{1}{z}\delta(z-z^{\prime})=\frac{1}{z_{0}^{2}}\sum_{p=1}^{\infty}%
J_{1}(k_{p}z)J_{1}(k_{p}z^{\prime}).
\end{equation}
Using this, the solution of \ref{A4} is then easily seen to be,%
\begin{equation}
f(k,z)=\sum_{p=1}^{\infty}\frac{1}{z_{0}^{2}c_{p}^{2}}\int_{0}^{z_{0}%
}dz^{\prime}\frac{J_{1}(k_{p}z)J_{1}(k_{p}z^{\prime})}{k^{2}-k_{p}^{2}}.
\end{equation}
Thus, one arrives at the following final form for the Green's function,%
\begin{equation}
G(z,x)=\frac{1}{z_{0}}\sum_{p=1}^{\infty}\alpha_{p}G_{p}(x)zJ_{1}(k_{p}z)
\label{Green}%
\end{equation}
where,%
\begin{align}
G_{p}(x)  &  =\int\frac{d^{4}k}{(2\pi)^{4}}\frac{e^{-ik\cdot x}}{k^{2}%
-k_{p}^{2}},\\
\alpha_{p}  &  =\frac{\int_{0}^{1}dxJ_{1}(x\chi_{p})}{\int_{0}^{1}dxxJ_{1}%
^{2}(x\chi_{p})}%
\end{align}

We shall now follow the procedure described by Itzykson and Zuber to find the
S-matrix that connects the fields before and after interaction with a
time-dependent source\cite{Zuber}. So imagine that the source, which obeys
$\partial_{\mu}J^{\mu}=0$, is turned on for a finite time $-T<t<T$. The "in"
and "out" fields, defined as $A_{in}^{\mu}=\lim_{t\rightarrow-\infty}A^{\mu}$
and $A_{out}^{\mu}=\lim_{t\rightarrow\infty}A^{\mu},$ are related by,
\begin{equation}
A_{out}^{\mu}(z,x)=A_{in}^{\mu}(z,x)+\int d^{4}x^{\prime}G^{-}(z,x-x^{\prime
})J^{\mu}(x^{\prime}),\text{ } \label{C1}%
\end{equation}
where $G^{-}\equiv G^{R}-G^{A}$ and $G^{R},G^{A}$ are, respectively, the
retarded and advanced Green's functions. $G^{-}$ is trivially obtained from
Eq.\ref{Green} in terms of the Green's functions for individual modes,
\begin{align}
G^{-}(z,x)  &  =\frac{1}{z_{0}}\sum_{p=1}^{\infty}\alpha_{p}G_{p}^{-}%
(x)zJ_{1}(k_{p}z),\\
G_{p}^{-}(x)  &  =G_{p}^{R}-G_{p}^{A}\nonumber\\
&  =i\int d\widetilde{k_{p}}\left(  e^{-ik\cdot x}-e^{ik\cdot x}\right)
\end{align}

The incoming and outgoing fields are also connected through a unitary operator
$S,$
\begin{equation}
A_{out}^{\mu}(z,x)=S^{-1}A_{in}^{\mu}(z,x)S,
\end{equation}
for which the following ansatz can be made,
\begin{equation}
S=e^{-i\int d^{4}xdzh(z)J_{\mu}(x)A^{\mu}(z,x)}%
\end{equation}
where $h(z)$ is as yet an unknown function. From the field commutation
relation in Eq.\ref{comm}, and from the Baker-Campbell-Haussdorf relation,
$e^{A}Be^{-A}=B+[A,B]$ (which holds for $[A,[A,B]]=0),$ it follows that,%
\begin{align}
A_{out}^{\mu}(z,x)  &  =S^{-1}A_{in}^{\mu}(z,x)S=A_{in}^{\mu}(z,x)+\label{C2}%
\\
&  \sum_{p=1}^{\infty}\int d^{4}x^{\prime}dz^{\prime}f(z^{\prime}%
)\frac{zz^{\prime}}{z_{0}^{2}c_{p}^{2}}J_{1}(k_{p}z)J_{1}(k_{p}z^{\prime
})G_{p}^{-}(x-x^{\prime})J^{\mu}(x^{\prime}).\qquad\nonumber
\end{align}
Setting equal the expressions for $A_{out}^{\mu}(z,x)$\ in Eq.\ref{C1} and
Eq.\ref{C2} forces the choice $h(z)=z^{-1}$ and leads to the important result,%
\begin{equation}
S=e^{-i\int d^{4}x\frac{dz}{z}J_{\mu}(x)A^{\mu}(z,x)}. \label{Smat}%
\end{equation}
From the asymptotic behaviour of $J_{1}(kz)$ for small $z$, it is clear that
the integrand above does not contain any singularity. \bigskip

From the S-matrix derived in Eq.\ref{Smat} above one can compute the amplitude
for the current $J_{\mu}(x)$\ to produce any number of vector mesons. Because
$A^{\mu}(z,x)$ in Eq.\ref{field} contains both creation $($inside $A_{\mu}%
^{-})$ and destruction $($inside $A_{\mu}^{+})$ operators, it is first
necessary to separate these by using the identity $e^{A+B}=e^{A}%
e^{B}e^{-[A,B]/2}$. This gives,
\begin{align}
S  &  =e^{-ig\int d^{4}x\frac{dz}{z}J^{\mu}(x)A_{\mu}^{-}(z,x)}e^{-ig\int
d^{4}x\frac{dz}{z}J^{\mu}(x)A_{\mu}^{+}(z,x)}\label{Sres}\\
&  \times e^{-g^{2}\sum_{p=1}^{\infty}\beta_{p}^{2}\int d\widetilde{k_{p}%
}\overrightarrow{J}^{\ast}(k)\cdot\overrightarrow{J}(k)}%
\end{align}
\qquad\qquad

where,
\begin{equation}
\beta_{p}^{2}=\frac{\left[  \int_{0}^{1}dxJ_{1}(x\chi_{p})\right]  ^{2}}%
{\int_{0}^{1}dxxJ_{1}^{2}(x\chi_{p})}, \label{beta}%
\end{equation}
and $\overrightarrow{J}(k)$ is the 4-d Fourier transform of $\overrightarrow
{J}(x),$%
\begin{equation}
\overrightarrow{J}(k)=\int d^{4}xe^{-ik\cdot x}\overrightarrow{J}(x).
\end{equation}
Note that $\overrightarrow{J^{\ast}}(k)=$ $\overrightarrow{J}(-k)$\ and that
only the physical polarizations have entered the calculations.

\ The probability for producing a single vector meson with polarization
$\lambda,$excitation $p$, and located in the momentum space element $d^{3}k$
is easily obtained from Eq.\ref{Sres} ,%
\begin{align}
dP  &  =\left\vert A\right\vert ^{2}d\widetilde{k_{p}}=\left\vert A\right\vert
^{2}\frac{d^{3}k}{(2\pi)^{3}2\omega_{p}}\label{prob1}\\
A  &  =-ig\beta_{p}\frac{\varepsilon(k,\lambda)\cdot J(k)}{Exp[g^{2}\sum
_{p=1}^{\infty}\beta_{p}^{2}\int d\widetilde{k_{p}}\overrightarrow{J}^{\ast
}(k)\cdot\overrightarrow{J}(k)]}. \label{prob2}%
\end{align}
The probability for emission of subsequent mesons, whether of the same type or
different, is uncorrelated with the first emission and is trivially obtained
from the above. Note that there is no delta function that conserves energy and
momentum in the final state. This follows from having assumed a heavy source
that does not suffer back reaction as it emits particles while moving on a
predetermined path.

\bigskip

\section{Synchrotron Radiation}

$\smallskip$What we have developed above is really a theory of bremsstrahlung
by a classical source coupled to quantum fields. The source, located in 4-d
spacetime, excites modes in the 5-d bulk that correspond to the excitation of
various vector meson states. In electrodynamics, the no-recoil assumption
limits the applicability of semi-classical bremsstrahlung theory to heavy
charged particles radiating soft zero-mass photons. But here, the lightest
particle that can be radiated has a mass around $770MeV/c^{2}$! So is there
any hope that vector meson bremsstrahlung can be observed?

The fundamental requirement of a non-recoiling source can possibly be met by a
large nucleus, such as $Au$, where the entire nucleus - rather than just
individual nucleons - couples to mesons. Indeed, coherent meson production
from nuclei by photons and other particles is a well-studied phenomenon. Let
us therefore consider a point source moving along a definite trajectory
$x(\tau)$ labelled by the proper time $\tau$. The current is,%
\begin{align}
J^{\mu}(t,\overrightarrow{y})  &  =\int d\tau\frac{dx^{\mu}}{d\tau}\delta
^{4}(y-x^{\mu}(\tau))\nonumber\\
&  =\int\frac{d^{4}k}{(2\pi)^{4}}e^{-ik\cdot y}J^{\mu}(k),
\end{align}
where,
\begin{equation}
J^{\mu}(k)=\int d\tau\frac{dx^{\mu}}{d\tau}e^{ik\cdot x(\tau)}. \label{cur}%
\end{equation}

Consider a heavy nucleus moving on a circular path in the $x-y$ plane with
radius $R$ and with frequency $\omega_{0}$. The coordinates of the particle
are $x^{\mu}=(t,R\cos\omega_{0}t,R\sin\omega_{0}t,0)$ implying that
$J_{z}(k)=0.$We choose axes such that $k^{\mu}=(\omega,k\sin\theta
,0,k\cos\theta).$ Using various Bessel identities it is straightforward to
show that,%
\begin{align}
J_{x}(k)  &  =\pi\omega_{0}R\sum_{n=n_{\min}}^{n=\infty}i^{n}\left[
J_{n+1}(kR\sin\theta)+J_{n-1}(kR\sin\theta)\right]  \delta(\omega-n\omega
_{0})\\
J_{y}(k)  &  =i\pi\omega_{0}R\sum_{n=n_{\min}}^{n=\infty}i^{n}\left[
J_{n+1}(kR\sin\theta)-J_{n-1}(kR\sin\theta)\right]  \delta(\omega-n\omega
_{0}).
\end{align}
Since $\omega^{2}=k^{2}+m_{p}^{2},$ $\omega$ has a minimum value equal to the
mass of the produced meson and so $n_{\min}=m_{p}/\omega_{0}=m_{p}R.$ This
reflects the fact that the agency which keeps the source in motion must pay
the price of creating the meson. \bigskip

Using Eqs.\ref{prob1}-\ref{prob2} let us work towards calculating the emission
probability, summed over final spins, to radiate a meson. This is proportional
to,%
\begin{align}
&  g^{2}\beta_{p}^{2}\frac{d^{3}k}{(2\pi)^{3}2\omega_{p}}\left(  \left\vert
J_{x}(k)\right\vert ^{2}+\left\vert J_{y}(k)\right\vert ^{2}\right)
\nonumber\\
&  =\delta(0)g^{2}\beta_{p}^{2}\frac{1}{4}\sum_{n=n_{\min}}^{n=\infty}\left[
J_{n+1}^{2}(kR\sin\theta)+J_{n-1}^{2}(kR\sin\theta)\right]  d\theta\sin\theta.
\end{align}
The $\delta(0)$ is a consequence of the fact that the source has been in
motion for an arbitrarily long time. It can be replaced by $t/(2\pi)$ thus
yielding a rate of emission proportional to,%
\begin{equation}
\frac{d^{2}P}{dtd\theta}\varpropto g^{2}\beta_{p}^{2}\frac{1}{8\pi}\sin
\theta\sum_{n=n_{\min}}^{n=\infty}k_{n}\left[  J_{n+1}^{2}(k_{n}R\sin
\theta)+J_{n-1}^{2}(k_{n}R\sin\theta)\right]  , \label{prob}%
\end{equation}
where,
\begin{align}
k_{n}  &  =\sqrt{n^{2}\omega_{0}^{2}-m_{p}^{2}}\\
\omega_{0}R  &  =\text{v}=\sqrt{1-\frac{1}{\gamma^{2}}}%
\end{align}
The proportionality constant in Eq.\ref{prob} is the square of the denominator
in Eqs.\ref{prob2}. The periodicity of the source motion implies that the
spectrum of the radiated particles is discrete$.$ Unfortunately there does not
seem to be a closed form for the series. However, one can readily check that
it is convergent provided for any finite $\gamma$ although the convergence
becomes increasingly slow as the source speed approaches that of light. The
series diverges for v$=c=1$. Since $n_{\min}=m_{p}/\omega_{0}$ is a large
number, the sum can be replaced by an integral,
\begin{equation}
\frac{d^{2}P}{dtd\theta}\varpropto\frac{g^{2}\beta_{p}^{2}}{4\pi}\sin
\theta\int_{\frac{m_{p}}{\omega_{0}}}^{\infty}dyk(y)J_{y}^{2}\left[
k(y)R\sin\theta\right]  , \label{rate}%
\end{equation}
with $k(y)=\sqrt{y^{2}\omega_{0}^{2}-m_{p}^{2}}.$

With this compact form, the emission rate for every member of the tower of
vector mesons can be estimated. The integral in Eq.\ref{rate} cannot be
performed in closed form nor by some straightforward numerical integration. To
obtain a rough estimate, we use Duhamel's formula for Bessel functions,
\begin{equation}
J_{n}(n\sin\alpha)=\frac{\left(  e^{\cos\alpha}\tan\frac{\alpha}{2}\right)
^{n}}{\left(  2n\pi\cos\alpha\right)  ^{1/2}},
\end{equation}
from which, at $\theta=\frac{\pi}{2},$ the rate from a single nucleus is
proportional to,%
\begin{equation}
\gamma^{3}e^{-\frac{2}{3}\gamma^{3}m_{p}R}.
\end{equation}
For a typical vector meson, and with $R\sim250$ metres, $m_{p}R\sim10^{18}. $
This requires $\gamma\sim10^{5}$ for reasonable emission rates. This is far
greater than the $\gamma$ of a heavy nucleus at RHIC, which is around
$1.5\times10^{2}$. Thus, the unfortunate conclusion is that vector meson
bremsstrahlung will be hard to detect in an accelerator ring. Nevertheless,
there are lessons to be learned here that will be useful in the next section.

\bigskip

\section{Ultraperipheral Collisions}

$\smallskip$Particle accelerators have large turning radii of the order of
kilometers so that charged orbiting particles can have low acceleration and
energy loss from bremsstrahlung is therefore minimized. From the point of view
of the formalism developed in this paper, this has the unfortunate implication
that meson bremsstrahlung is strongly suppressed. To test our ideas we shall
now turn to the ultraperipheral high-energy collision (UPC) of two heavy ions.
In an UPC the two ions interact electromagnetically rather than hadronically,
requiring that the impact parameter $b>2R$. After colliding and producing a
$\rho^{0}$\ the colliding nuclei can remain in the ground state, or perhaps
transit to an excited state. For either case, the STAR collaboration has
recently measured $\rho^{0}$ and direct $\pi^{+}\pi^{-}$ production in Au-Au
collisions at $\sqrt{s}=200GeV$/nucleon collisions at RHIC\cite{star}. UPCs
are part of the heavy ion program at ALICE, ATLAS, and CMS at CERN. For a
review, the reader is referred to refs\cite{UPC1},\cite{UPC2}.

In the normal QCD analysis, the colliding nuclei in a UPC are the source of an
intense pulse of photons, the equivalent photon flux being determined from the
Fourier transform of the electromagnetic field of the moving charges. These
photons produce various mesons from elementary photon-photon and
photon-pomeron vertices. Rather than individual nucleons, the entire nucleus
produces the photon and pomeron flux, i.e. the nucleons act coherently and
cooperatively without betraying the internal nuclear structure. Our picture of
meson production will be apparently very different, but in fact it will be
fairly similar. Each nucleus is the means for providing acceleration to the
other through Coulomb repulsion. Moreover, since the entire nucleus turns
without breaking up or excitation, it can be considered as a point particle.
Because the "turning radii" are over nuclear length scales rather than the
macroscopic scales, the accelerations can be sufficiently large to cause
copious meson emission.

Consider, therefore, two identical ultrarelativistic point charges moving
towards each other and then scattering through a small angle $\theta$. Both
have charge $Ze,$ mass $M,$ and four-velocities v$_{1}^{\mu}$and v$_{2}^{\mu}%
$. The transverse separation between the charges (impact parameter) is $b.$
The classical trajectory for non-relativistic charges is, of course,
hyperbolic. This undergoes modification in the relativistic case. However,
even for non-releativistic motion, the integrals needed for calculating the
$k$-space current are formidably difficult. We shall, therefore, use a
caricature of the actual classical path by demanding that the charges collide
at proper time $\tau=0$ after which they suddenly change their (constant)
four-velocities from v$_{1}^{\mu}$ and v$_{2}^{\mu}$\ \ to v$_{1}^{\prime\mu}%
$and v$_{2}^{\prime\mu}$ respectively,
\begin{align}
x_{1}^{\mu}  &  =\text{v}_{1}^{\mu}\tau+\frac{1}{2}b^{\mu}\text{ \ }%
\ \ \tau<0\\
x_{1}^{\prime\mu}  &  =\text{v}_{1}^{\prime\mu}\tau+\frac{1}{2}b^{\mu}\text{
\ \ }\tau>0\\
x_{2}^{\mu}  &  =\text{v}_{2}^{\mu}\tau-\frac{1}{2}b^{\mu}\text{ \ \ \ }%
\tau<0\\
x_{2}^{\prime\mu}  &  =\text{v}_{2}^{\prime\mu}\tau-\frac{1}{2}b^{\mu}\text{
\ \ }\tau>0
\end{align}
The current in Fourier space follows from Eq.\ref{cur},%
\begin{align}
J^{\mu}(k)  &  =J_{1}^{\mu}(k)+J_{2}^{\mu}(k)\nonumber\\
&  =i\left(  \frac{\text{v}_{1}^{\prime\mu}}{k\cdot\text{v}_{1}^{\prime}%
}-\frac{\text{v}_{1}^{\mu}}{k\cdot\text{v}_{1}}\right)  e^{\frac{i}{2}k\cdot
b}+i\left(  \frac{\text{v}_{2}^{\prime\mu}}{k\cdot\text{v}_{2}^{\prime}}%
-\frac{\text{v}_{2}^{\mu}}{k\cdot\text{v}_{2}}\right)  e^{-\frac{i}{2}k\cdot
b}. \label{J}%
\end{align}
Only the square of the 3-vector $\overrightarrow{J}$ needs to be computed.
This has direct terms corresponding to vector meson emission from each nucleus
separately, as well as an interference term corresponding to simultaneous
emission from both nuclei,%
\begin{equation}
\left\vert J\right\vert ^{2}=\left\vert \overrightarrow{J}_{1}\right\vert
^{2}+\left\vert \overrightarrow{J}_{2}\right\vert ^{2}+2\cos(\overrightarrow
{k}\cdot\overrightarrow{b})\left\vert \overrightarrow{J}_{1}\right\vert
\left\vert \overrightarrow{J}_{2}\right\vert .
\end{equation}
With $\hat{x}\ $and $\hat{z}\ $denoting unit vectors as usual, we now make a
definite choice of velocity vectors:
\begin{align}
\text{v}_{1}^{\mu}  &  =\gamma(1,\hat{z}V)\text{ \ \ \ \ \ \ v}_{1}^{\prime
\mu}=\gamma(1,\,\hat{z}V\cos\theta+\hat{x}V\,\sin\theta)\\
\text{v}_{2}^{\mu}  &  =\gamma(1,-\hat{z}V)\text{ \ \ \ \ v}_{2}^{\prime\mu
}=\gamma(1,-\hat{z}V\,\cos\theta-\hat{x}V\,\sin\theta)
\end{align}
$\ $For ultrarelativistic nuclei, $V\approx1.$ As before,$\ k^{\mu}%
=(\omega,\overrightarrow{k})$ with $\omega^{2}=k^{2}+m^{2}$ and $b^{\mu
}=b(0,\hat{x})$.\ \ For an UPC, the nuclei undergo scattering through very
small angles only. With $\theta<<1,$ a compact form results for $\left\vert
J\right\vert ^{2}.$ The direct term is,%
\begin{equation}
\theta^{2}\left\{  \frac{1}{(\omega-k_{z})^{2}}+\frac{1}{(\omega+k_{z})^{2}%
}+\frac{k_{x}^{2}}{(\omega-k_{z})^{4}}+\frac{k_{x}^{2}}{(\omega+k_{z})^{4}%
}\right\}  ,\nonumber
\end{equation}
and the interference term is,%
\begin{equation}
-\frac{2\theta^{2}}{(\omega^{2}-k_{z}^{2})}\left\{  1+\frac{k_{x}^{2}}%
{(\omega^{2}-k_{z}^{2})}\right\}  \cos(k_{x}b).
\end{equation}
Expressed in terms of the rapidity variable $y$,%
\begin{equation}
y=\frac{1}{2}\log\frac{\omega+k_{z}}{\omega-k_{z}}%
\end{equation}%
\begin{equation}
\left\vert J\right\vert ^{2}=2\theta^{2}\left\{  \frac{m^{2}(\cosh2y-\cos
k_{x}b)+k_{x}^{2}(\cosh4y+\cosh2y)}{(m^{2}+k_{x}^{2})^{2}}\right\}  .
\end{equation}

From Eq.\ref{J} or the subsequent results, we see that $\left\vert
J\right\vert ^{2}\propto1/k^{2}$ and hence the total crossection is
logarithmically divergent at the upper momentum limit. This is a consequence
of the discontinuous change in the velocity; a continuous hyperbolic path
would not suffer from this problem. Indeed, we can see that the circular
motion case would lead to finite crossections.

The scattering angle $\theta$ is determined by the impact parameter $b$, as
can be seen from a simple calculation using the retarded electric field of a
relativistic charge that passes by a second similar charge\cite{JDJackson}.
Assuming that neither trajectory deviates appreciably from a straight line,
the transverse momentum impulse is,
\begin{equation}
\Delta p_{x}=2\frac{Z^{2}e^{2}}{b},
\end{equation}
and hence the scattering angle is,%
\begin{equation}
\theta=\frac{\Delta p_{x}}{p}=\frac{2Z^{2}e^{2}}{\gamma M}\frac{1}{b}.
\end{equation}
The kinetic energy of the non-relativistic transverse motion is,%
\begin{equation}
\Delta E=2\times\frac{\left(  \Delta p_{x}\right)  ^{2}}{2M}=\frac{4Z^{4}%
e^{4}}{M}\frac{1}{b^{2}},
\end{equation}
which, for small enough $b$, could provide sufficient energy for particle production.

The crossection for meson production is easily computed because, having
started from the premise that there is no back-reaction on the emitting
source, it is clear that there are no complicated phase space factors. This
limits the validity of our approach to low meson momenta. In fact, application
to pion production would be more justifiable than to vector meson production.
However, as we shall soon see, there seems to be fair agreement with data even
for $\rho^{0}$ production.

Since $g,$\ i.e. the coupling of vector mesons to the source, is unknown, it
is sufficient to write proportionality relations. As a first step, note that
the number of nuclei scattered per unit time (v$\approx c=1$) around angle
$\theta$\ is,%

\begin{equation}
dN\propto2\pi bdb\propto\frac{d\theta}{\theta^{3}}.
\end{equation}
This is identical to the Rutherford (or Mott) crossection behaviour in the
forward direction. Multiplication by the emission probability yields,%
\begin{equation}
d\sigma\propto\beta_{p}^{2}\left\vert J\right\vert ^{2}d\widetilde{k_{p}}%
\frac{d\theta}{\theta^{3}}.
\end{equation}
Integrating over $\theta$ or, equivalently, over $b,$ yields the crossection
in the rapidity variable $y$ and the transverse momentum $k_{\perp}=k_{x}$,%
\begin{equation}
d\sigma\propto\frac{\beta_{p}^{2}}{m_{p}^{2}}\int_{b_{\min}}^{b_{\max}}%
\frac{db}{b}\frac{\cosh2y-\cos k_{\perp}b+\frac{k_{\perp}^{2}}{m_{p}^{2}%
}(\cosh4y+\cosh2y)}{(1+\frac{k_{\perp}^{2}}{m_{p}^{2}})^{2}}d\widetilde{k_{p}%
}.
\end{equation}
An easy integration gives,

\bigskip%
\begin{equation}
\frac{d\sigma}{dyd^{2}k_{\perp}}\propto\frac{\beta_{p}^{2}}{m_{p}^{2}%
(1+\frac{k_{\perp}^{2}}{m_{p}^{2}})^{2}}\left\{
\begin{array}
[c]{c}%
\left[  \cosh2y+\frac{k_{\perp}^{2}}{m_{p}^{2}}(\cosh4y+\cosh2y)\right] \\
\times\log\frac{b_{\max}}{b_{\min}}-\operatorname{Ci}(k_{\perp}b_{\max
})+\operatorname{Ci}(k_{\perp}b_{\min})
\end{array}
\right\}
\end{equation}
where $\operatorname{Ci}$ is the standard cosine integral. The lower limit,
$b_{\min}=2R\approx14$ fm$,$ corresponds to the nuclei just touching each
other, while the upper limit is determined by requiring that the scattering be
sufficiently hard so as to produce at least one meson, $b_{\max}^{2}%
\approx\frac{4Z^{4}e^{4}}{Mm_{p}}.$The values of $\beta_{p}^{2}$\ (see
Eq.\ref{beta})decrease steadily with $p$: $\beta_{1}^{2}=1.28,$ $\beta_{2}%
^{2}=0.57,$ $\beta_{3}^{2}=0.36$ showing that higher resonances will be
produced in lesser amounts.%

\begin{figure}
[ptb]
\begin{center}
\includegraphics[
height=2.9784in,
width=4.849in
]%
{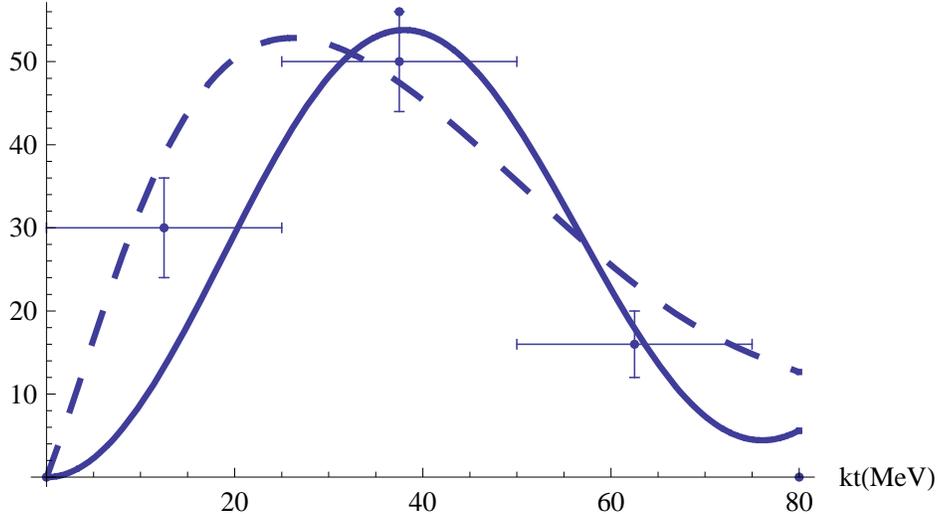}%
\caption{The differential crossection $\frac{d\sigma}{dyd^{2}k_{\perp}}$for
y=0 calculated in ADS as a function of transverse meson momentum $k_{\perp}$
(solid line) compared against the photon-pomeron fusion calculation of Hencken
et al (dashed line)\cite{transverse}. . The vertical scale is chosen
arbitrarily as normalizations cannot be calculated in our model. Also shown
are data points from the Star Collaboration\cite{star} for the number of
$\rho^{0}$ counts, binned in 25 MeV intervals.}%
\end{center}
\end{figure}

In fig.1 the crossection, arbitrarily normalized, is plotted for $\rho^{0}$
production in Au-Au UPCs as a function of transverse momentum for $y=0$ and
compared against an existing calculation based on photon-pomeron fusion
\cite{transverse}. Also shown are data points from the Star
Collaboration\cite{star} for the number of $\rho^{0}$ counts, binned in 25 MeV
intervals. Since $g$ is not known, absolute magnitudes cannot be predicted in
the ADS model. However, the shape of the momentum distribution is not
dissimilar from either experiment or conventional theory. The interference
term is crucial, as was found earlier in ref \cite{transverse}.

\bigskip In summary, we have calculated the quantum fluctuations induced in
the 5-D bulk when a point source coupled to vector fields in 4-D space-time is
transported along a classical trajectory. The quantum fluctuations amount to
the production of different mesons with different momenta in 4-D. Meson
crossections are calculable as a function of the point source's motion. In
principle, the motion of a heavy nucleus in an accelerator could lead to the
emission of massive particles similar to photon bremsstrahlung but, in
practice, the rate is extremely small unless the nuclei have extremely large
gamma-factors. On the other hand, for the ultraperipheral collisions of heavy
ions, the rates are appreciable. The ADS formalism allows for the prediction
of the transverse momentum spectrum. The comparison with existing conventional
calculations is fairly satisfactory, and broad features of the existing data
are reproduced reasonably well. Emission rates for various excited meson
states can be predicted with no additional parameters.

$\smallskip$

{\Large Acknowledgments\medskip}

The author thanks Tom Cohen for many enjoyable conversations and for sharing
critical insights. He would also like to thank other members of the TQHN group
at the University of Maryland, particularly Xiangdong Ji and Steve Wallace,
for gracious hospitality during a visit made in July 2008 when part of this
work was done.


\begin{thebibliography}{99}                                                                                               %


\bibitem {Theory}J. Maldacena, \textquotedblleft The Large N limit of
superconformal field theories and supergravity\textquotedblright, Adv.Theor.
Math. Phys. 2:231, 1998, hep-th/9711200; E. Witten, \textquotedblleft Anti-de
Sitter space and holography\textquotedblright, Adv. Theor. Math. Phys.2: 253,
1998, hep-th/9805028; L. Susskind and E. Witten, \textquotedblleft The
Holographic Bound in Anti-de Sitter Space\textquotedblright, hep-th/9805114

\bibitem {DIS}J. Polchinski and M. Strassler, "Hard scattering and gauge-
string duality". Phys.Rev.Lett.88:031601,2002, hep-th/0109174; J. Polchinski
and M. Strassler, JHEP 0305:012,2003, hep-th/020921; S. J. Brodsky and G. F.
de Teramond, "Light-front hadron dynamics and AdS/CFT correspondence," Phys.
Lett. B 582, 211 (2004) [arXiv:hep-th/0310227].

\bibitem {Spectrum}H. Boschi-Filho and N. R. F. Braga, "Gauge string duality
and scalar glueball mass ratios," JHEP 0305, 009 (2003)
[arXiv:hep-th/0212207]; S.J.Brodsky and and G.F.de Teramond, "Hadronic spectra
and light-front wavefunctions in holographic QCD",
Phys.Rev.Lett.96:201601,2006, e-Print: hep-ph/0602252.

\bibitem {Son}J. Erlich, E. Katz, D. T. Son and M. A. Stephanov, "QCD and a
holographic model of hadrons," Phys.Rev.Lett. 95, 261602 (2005) [arXiv:hep-ph/0501128].

\bibitem {Rho}S. Hong, S. Yoon and M. J. Strassler, "On the couplings of
vector mesons in AdS/QCD," JHEP 0604, 003 (2006) [arXiv:hep-th/0409118]; "On
the couplings of the rho meson in AdS/QCD," hep-ph/0501197.

\bibitem {FF}H. J. Kwee and R. F. Lebed, "Pion Form Factors in Holographic
QCD," JHEP 0801, 027 (2008) [arXiv:0708.4054[hep-ph]]; S.J.Brodsky and and
G.F.de Teramond, "Light-Front Dynamics and AdS/QCD Correspondence:
Gravitational Form Factors of Composite Hadrons", Phys.Rev.D78:025032,2008,
arXiv:0804.0452 [hep-ph]; H. R. Grigoryan and A. V. Radyushkin, "Pion Form
Factor in Chiral Limit of Hard-Wall AdS/QCD Model," Phys. Rev. D 76, 115007
(2007) [arXiv:0709.0500 [hepph]]; H. R. Grigoryan and A. V. Radyushkin, "Form
Factors and Wave Functions of Vector Mesons in Holographic QCD," Phys. Lett. B
650, 421 (2007) [arXiv:hep-ph 0703069]. H. J. Kwee and R. F. Lebed,
\textquotedblleft Pion Form Factors in Holographic QCD,\textquotedblright%
\ arXiv:0708.4054 [hep-ph]; D. Rodriguez-Gomez and J. Ward, \textquotedblleft
Electromagnetic form factors from the fifth dimension,\textquotedblright%
\ arXiv:0803.3475 [hep-th].

\bibitem {GPD}Z.Abidin and C.Carlson, "Gravitational form factors of vector
mesons in an AdS/QCD model", Phys.Rev.D77:095007,2008, hep-ph 08013839

\bibitem {Kaon}T. Hambye, B. Hassanain, J. March-Russell and M. Schvellinger,
"On the Delta(I) = 1/2 rule in holographic QCD," Phys. Rev. D 74, 026003
(2006) [hep-ph/0512089]; "Four-point functions and kaon decays in a minimal
AdS/QCD model," Phys. Rev. D76, 125017 (2007) [hep-ph/0612010].

\bibitem {Cohen}T.Cohen, "Challenges facing holographic models of QCD",
arXiv:hep-ph 08054813.

\bibitem {Braga}H. Boschi-Filho and N.Braga, "Bulk versus boundary quantum
states", Phys.Lett.B525:164-168,2002, e-Print: hep-th/0106108.

\bibitem {Zuber}C.Itzykson and J.Zuber, "Quantum Field Theory", McGraw-Hill
International Book Co. (1980)

\bibitem {star}Abelev et al, "$\rho^{0}$ Photoproduction in Ultra-Peripheral
Relativistic Heavy Ion Collisions with STAR.", Star Collaboration,
Phys.Rev.C77:34910,2008. E-Print: arXiv:0712.3320

\bibitem {UPC1}C.A. Bertulani, S.R. Klein, J.Nystrand, "Physics of
ultra-peripheral nuclear collisions.", Ann.Rev.Nucl.Part.Sci.55:271-310,2005,
E-Print: nucl-ex/0502005. \ 

\bibitem {UPC2}K.Hencken, "The Physics of Ultraperipheral Collisions at the
LHC", Phys.Rept.458:1-171,2008, e-Print: arXiv:0706.3356

\bibitem {JDJackson}J.D.Jackson, Classical Electrodynamics, second edition,
John Wiley 1975.

\bibitem {transverse}K. Hencken, G. Baur, D. Trautmann, "Transverse momentum
distribution of vector mesons produced in ultraperipheral relativistic heavy
ion collisions", Phys.Rev.Lett.96:012303,2006, e-Print: hep-ph/0506014.
\end{thebibliography}
\end{document}